\documentclass[review]{elsarticle}
\usepackage{lineno,hyperref}
\usepackage[english]{babel}
\usepackage{graphicx}
\usepackage{amsmath}
\graphicspath{ {./jpg/} }
\modulolinenumbers[5]
\journal{Future Generation Computer Systems}

\begin{document}

\begin{frontmatter}

\title{Interactive distributed cloud-based web-server systems for the smart healthcare industry}

\author{Kondybayeva Almagul Baurzhanovna\fnref{myfootnote}}

\fntext[myfootnote]{post-graduate in a National University of Science and Technology "MISiS", the Cybernetics department, Russian Federation, Moscow, 4, Leninsky ave}

\address[mymainaddress]{alma.kond@gmail.com}
\address {National University of Science and Technology MISiS}

\begin{abstract}

\paragraph {Subject} This work is dedicated to the questions of the contemporary medical image visualization, the architecture design of the cloud server systems and the using of methods for the .DICOM data representation for the distributed smart healthcare industry systems.

\paragraph {Purpose} In modern medicine and biology, in the pace of research along with the objective need for a constant increase, there is a sharp necessity of the three-dimensional data representation \cite{Tractica}: requiring high-performance methods of the three-dimensional visualization, processing, decomposition, reconstruction and analysis.
In many ways, the reason for the sharp growth is associated with the development of the computer technologies (from quantity into quality): the quantitative characteristics of the devices and computers used, the emergence of three-dimensional technologies for processing, visualization and research of data, and, on this basis, the creation of new three-dimensional methods for the human-machine interaction.

\paragraph {Research methodology} This paper proposes a method for  visualization using direct volumetric rendering based on the cubic spline interpolation on a ray emission technology. Modifications of these algorithms, based on a block decomposition, are proposed and still investigated. The server architecture is proposed as a cloud hypervisors server system for the grid processing of the data required.

\paragraph {Research results} The well-known method of block decomposition of the medical data was studied: the idea of block searching was implemented, auxiliary structures reducing the volume range were used to skip empty spaces (empty space leaping). The cloud architecture for the server processing was developed as well as the user interface was proposed.

\paragraph {Scope of application} The work aims to investigate the possible contemporary interactive cloud based solutions in the fields of the applied medicine for the smart Healthcare as the data visualization open-source free system distributed under the MIT license. 

\paragraph {Conclusions} 
A comparative study of a number of the well-known implementations of the Ray Casting algorithms was studied. A new method of numerical calculus is proposed for calculating the volume -- the method of spheres, as well as a proposal for paralleling the algorithm on graphic accelerators in a linearly homogeneous computing environment using the block decomposition methods. For the artifacts control -- algorithm of the cubic interpolation was used. The cloud server architecture was proposed. The work is done as a part of the PhD thesis of the author for non-profit/non-commercial, educational/research only reasons under the MIT License. 
\end{abstract}

\begin{keyword}
\texttt{Data Science for Smart Healthcare}\sep healthcare \sep internet of things \sep computer science \sep interactive visualization \sep distributed \sep web-server \sep cloud \sep parallel \sep systems \sep scientific research \sep cloud technologies \sep communications \sep information \sep cloud healthcare systems \sep web-services \sep web-technologies \sep information systems \sep experimental research \sep DICOM \sep CT \sep computed tomography \sep medical imaging \sep volume render
\MSC[2020] 68-00\sep  97U99 
\end{keyword}

\end{frontmatter}


\section{Introduction}

\paragraph{Introduction} 
One of the modern highly informative methods in medical diagnostics
is the \textit{computed tomography} technology (CT).
\textit{X-ray computed tomography} or CT is an imaging process that reproduces cross-sectional image representing the X-ray attenuation properties of the body. 
Three-dimensional reconstruction of CT image slices allows to visualize in three-dimensional space the localization of blood vessels, pathologies and other features, making this technique useful as an interactive visualization tool for the Healthcare purposes. In present days, the duration of the whole body computed tomography (with a slice layer thickness of less than $ \leq$ 1 mm) is about $ \sim$ 10-15 seconds, and the result of the study is from several hundred up to several thousand images \cite{Calhoun}. Actually modern \textit{Multislice CT} (MSCT) is a method for volumetric studies of the entire human body (due to the the resulting axial transverse tomography constitute a three-dimensional data array) allowing to perform any reconstruction of images, including multi-plane reformation, volumetric (if necessary, stereo) visualization, virtual endoscopy \cite{Ding:2697996}.

\begin{figure}[h]
\centering
\includegraphics[scale=0.35,angle=0]{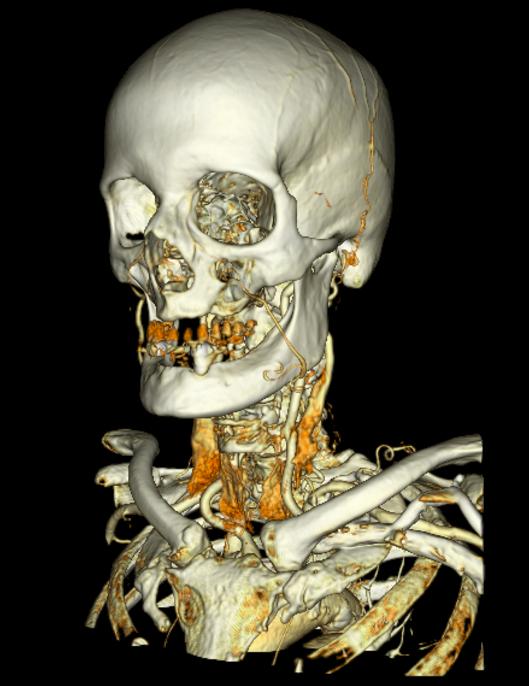}
\caption{The 3D volume render example made with the cubic spline interpolation}
\centering
\end{figure}

\paragraph{Aims of the study}
The object of research is spatial data science in the fields of the medicine and proper healthcare industry in computed tomography (CT) of existing types in .DICOM format. The subject of the study are methods and algorithms of the high-quality visualization of medical spatial data, especially medical image tomography data, as well as methods and algorithms for morphological representation of surfaces and features on GPU/CPU systems including server side cloud system architecture with the following steps:
\begin{enumerate}[(a)]
\item to create the sustainable model of the homogeneous cloud based server architecture,
\item to create the reliable model of the web-server,
\item to create the representation for the volume data render tasks (including mobile operating systems).
\end{enumerate}
\textit{The objects of study} are the models of synthesis systems' images for the three-dimensional models of human-machine communication in real time in the analysis of medical and biological spatial data, as well as the development and the study of models and architectures for the GRID computing and the cloud server hypervisor implementations. This goal requires the following tasks: 
\begin{itemize}
	\item to study the existing methods of the three-dimensional visualization in applied medicine, and science in general, an analysis of approaches for improving the quality and productivity of the 3D renderings,
	\item to develop existing methods for their complementary application in conditions of budgetary GPUs and the creation of 3D visualization technology for mass medical applications, including: 
	\begin{enumerate}[(a)]
		\item modification of the method of the block decomposition of gigavoxel (more than $\sim$\ $10 ^ {9}$\ voxels) data for visualization on the GPU, preserving the possibilities and quality of visualization for not decomposed data (the possibility of using cubic visual interpolation, lighting, casting shadows, skipping blank areas, different integration conditions along the beam, ...),
		
		\item to develop methods to improve the performance of 3D visualization and suppression in the various kinds of render defects (artifacts) rational in GPU conditions,
		\item to develop a method for quantifying visualization quality for
		achieving the required quality and comparing the real (in compliance with a given
		quality) the effectiveness of the proposed methods.
	\end{enumerate}
	\item to study and develop an online prototype of the software package that implements the proposed methods on	modern parallel hardware GPU architectures, and	experimentally investigate the effectiveness of methods.
\end{itemize}

\paragraph{Render example} The Figure 1 represents the 3D volume render data visualization made with the cubic spline interpolation used in the study which is considered to be used as a basic algorithm for the volume render process in this system.
\section{Methodology}
There are two main ways to visualize scalar field isosurfaces:
\begin {enumerate} [(a)] 
\item to restore isosurface geometry (usually in polygonal form). In this case, the well-known marching cubes method is usually used. For surfaces reconstructed by this method, there is an algorithm for their effective compression. The main advantage of this approach is low hardware requirements: in contrast to the method of emitting rays, the process of rendering a polygon surface does not impose high requirements on the GPU, in addition, there is no need to store the initial field data for visualizing the surface. Another advantage of the approach is access to polygons, for example, for analysis of surface morphology. About the disadvantages, the duration of restoration of isosurface geometry can be noted \cite{Flohr},
\item visualization using the method of emitting rays (the Ray Casting method) does not imply the restoration and preservation of the surface mesh in the RAM (the Random Access Memory). The method is to emit a beam for each pixel in the image in order to find collisions of the beam with the surface and determining the illumination of the surface at a given point, and thus determining the color of the pixel. Obviously, generating an image of an isosurface by this method is a much more resource-intensive process than rendering a polygonal surface, so the program to find collisions are usually performed on the GPU. In this case, the advantage is the lack of the need to restore polygons of the surface, allowing interactive adjustment of the value for the isosurface. Often, on modern video cards, the method of emitting rays is superior to the usual rendering of the polygonal mesh, both in quality and in performance. However, the need to store the original data array and high hardware requirements limit the use of this method \cite{Hadwiger}.
\end{enumerate}
The ray tracing method and its modification, called the method of the emission of the rays (Ray Casting -- see Figure 2) allows to achieve the best quality and informative volume rendering.
In this method, the color of each pixel of the desired image is calculated and the corresponding beam then is generated; which is a point in the space (for example, the position of the observer) and the direction of the ray.
Moving in this direction with a certain step, the beam “accumulates” the color of the pixel.
An important advantage of the ray emission method is that the algorithm is easy
paralleling on the graphics processing unit (GPU) since each pixel of the desired image is processed independently of the rest. Medical image size \textit{512*512*512} voxels can easily fit in GPU memory and the most efficient storage for such data is a three-dimensional texture; since the GPU provides texture access caching and automatic trilinear interpolation of data when sampling at an arbitrary point in space \cite{Engel}. In this work the cubic interpolation is being used.
The following method is proposed in this study: modification of the method of block decomposition of data in the Ray Casting algorithm -- different optimal sequence bypass blocks using volume optimized auxiliary structures of the method skipping empty areas (empty space leaping), providing the ability to build local lighting and shadows combined with tables of the previously integrated rendering. 
The task of three-dimensional visualization of scalar and vector fields in medical and scientific imaging in general: visualization of scalar fields defined on regular and irregular grids.
\begin{figure}[h]
	\centering
	\includegraphics[scale=0.75,angle=0]{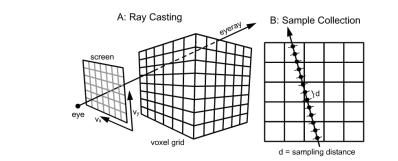}
	\caption{The Ray Casting algorithm}
	\centering
\end{figure}
\paragraph{Voxel graphics}The method is based on the display process of three-dimensional textures in space. Naturally, texture elements (voxels) are usually translucent. As a rule, scalar fields are visualized in this way. The method of direct volumetric rendering, which is largely devoted to this work, is intended to visualize voxels as a translucent medium \cite{Hernell}.
\paragraph{Isosurfaces} Isosurfaces are intended for visualization of three-dimensional scalar fields. Sometimes several isosurfaces are displayed at once; their color may depend on the field value \cite{Donatelli:2704061}.
Three-dimensional texture is an acceptable repository for volumetric data rendering performed on graphic video cards using OpenGL library extensions. However, there are restrictions on the size of the texture: the maximum size is 5123 voxels \cite{Calhoun} \cite{Ding:2697996}. Using a block view of the data circumvents this limitation. For programming on the GPU, the shader language GLSL was chosen, because at the moment, the performance of implementations on CUDA and, especially, OpenCL is often inferior to shader implementations \cite{Dobashi:2710749}. The main limitation of shaders is the lack of access to shared memory on the GPU \cite{Calhoun}. This access allows you to group the rays into packets (slabs). Using Slab-based rendering allows decomposition in the image space, grouping the rays into packets with a shared shared memory, which cannot be accessed from shaders \cite{MargretAnouncia:2697932}. When decomposing data, different data blocks are written into different three-dimensional textures. Textures are the same size, with the exception of those that capture only part of the source data. Using blocks of different sizes and covering only the visible part of the data can give an additional performance boost and save GPU memory. The texture map method also reduces the size of the used GPU memory \cite{Calhoun}. In this implementation, there is a limitation to excluding completely empty blocks and fitting the bounding box for each block to the visible voxels. To avoid artifacts at the junction, adjacent blocks must overlap at least one voxel thick. This will be enough if trilinear interpolation is used in the rendering algorithm when sampling values ​​from the data. In this implementation, the blocks overlap by a thickness of three voxels is used. It is used due to the reasons that, firstly, for the Fong local lighting model \cite{Donchin}, to calculate the gradient, for which it is necessary to make additional samples from neighboring voxels, and secondly, to use cubic interpolation during the sampling instead of trilinear, including calculating the gradient, which extends the radius of the sample by one more voxel. There are also approaches for stitching data blocks having different spatial resolutions.
\begin{figure}[h]
	\centering
	\includegraphics[scale=0.25,angle=0]{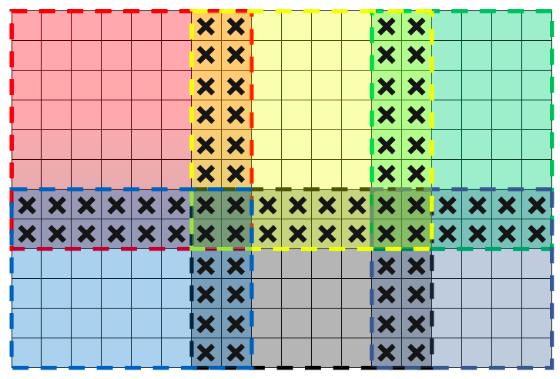}
	\caption{Decomposing data into overlapping blocks structure}
	\centering
\end{figure}
Figure 3 represents decomposing data into overlapping blocks. Here are the blocks overlap by a thickness of two voxels. Blocks marked in red and yellow have regular size (8 voxels in this case). The remaining blocks capture the remains data and are reduced in size if one wants to visualize three-dimensional discrete data array (tomogram) by volume rendering method, then to each possible data value one must set certain optical properties. Transfer function \textit{T(x)} (transfer function), or palette, in the simplest case, matches any color value and transparency (usually in practice, the value is stored in memory opacity, and \textit{0 - full transparency, 1 - maximum opacity)}.
\textit{Postclassification} - the operation scheme of the basis algorithm on
render in which coloring (classification) of a point in space occurs
after interpolating the value selected at the current point of the beam, i.e. color is determined as the interpolated data value.
\textit{Preclassification} - the operation scheme of the volume rendering in which the color of a space point is defined as color interpolation nearby voxels that are classified (as received color) according to their values, i.e. coloring occurs before interpolation \cite{Calhoun}.
If one visualizes the array as a lot of multi-colored translucent cubes, i.e. if interpolation between data cells (voxels) is not used, then the result of the 3D rendering will contain crude artifacts. As for \textit{“smoothness”} of visualization it is necessary to use interpolation between the nodes of the source (data) grid; and with it two different approaches may arise. In the most widespread cases the postclassification method is being used \cite{Calhoun} \cite{Tarini}. 

The optical properties of a point in space is first calculating the interpolating value of \textit{V} data at a point (usually using trilinear interpolation \cite{Calhoun}), the value of the transfer function is considered as point \textit{V}, i.e. \textit{T(V)} even if the interpolated value of \textit{V} does not occur at all in the histogram of the source data \cite.
On the contrary, in preclassification process, all voxels are “painted” at first, but beyond
optical properties, the arbitrary point in space is considered as the result of the interpolation between the optical properties (color with transparency) of voxels, i.e. classification occurs before interpolation, not after \cite{Hernell} \cite{Hadwiger} \cite{Tarini} \cite{Sattler}.
In practice, a model with postclassification is more often used, because it gives:
\begin{enumerate}[(a)]
	\item significantly improving visualization quality: significantly less noticeable stepping data comparing to the preclassification with tricubic interpolation for rendering (preclassification gives the same artifacts),
	\item better performance or better resource intensity: on practice for preclassification method each voxel and its optical properties must be either stored in the memory, or calculate these voxels' properties during rendering. 
\end{enumerate}
In the first case instead of a 12-bit array with the source data in the GPU one shall need to load a 32-bit array of the same dimensions, which will store the color and voxel transparency instead of the original data values. And when changing the transfer function it is needed to re-calculate and load the entire array, instead of loading a new transfer function. In the second case one has to make samples of eight voxels (in the case of trilinear interpolation) calculate their colors and transparency and then find the interpolated value of color and transparency for the point. Whereas in postclassification one makes one sampling from a point using trilinear filtering of textures, which practically gives hardware for free
trilinear interpolation  \cite{Shanmugam} \cite{Stewart} \cite{Besset:768962}.
\begin{figure}[h]
	\centering
	\includegraphics[scale=0.25,angle=0]{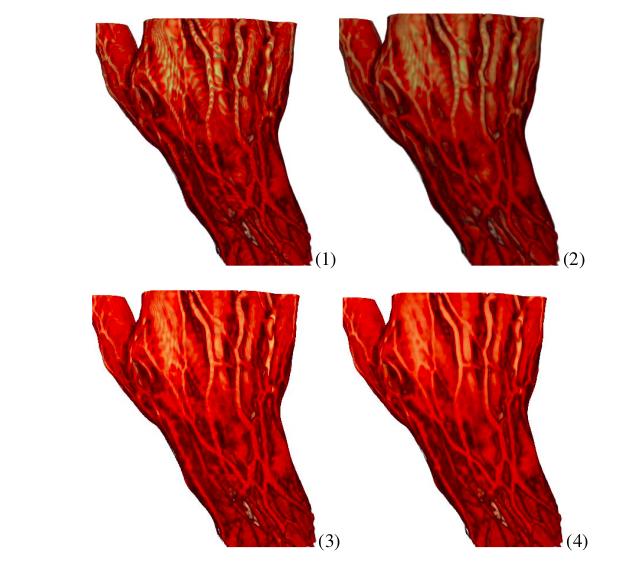}
	\caption{Comparison of rendering results: 1) preclassification +
		trilinear interpolation; 2) preclassification + tricubic interpolation; 3)
		postclassification + trilinear interpolation; 4) postclassification + tricubic
		interpolation.}
	\centering
\end{figure}
In this work the preference is given to the tricubic spline interpolation using postclassification algorithm due to the results from the comparison presented on the Figure 4.
\paragraph{Analysis} 
Analysis of the problems of implementing the method of block decomposition under conditions of maintaining high quality 3D visualization.
\paragraph{Pros}:
\begin{enumerate}[(a)]
	\item the ability to load large amounts of data. In addition, when divided into blocks of 643 voxels, one can save the blocks that do not contain useful information. For example, for CT scans, one can usually drop $ \sim$ 40 $ \%$ of the blocks, due to the reason that these blocks contain only air. The size of the visualized data is limited only by the capacity of the video card (in the case of a consumer video card with a memory of only 1 GB can visualize an array of data up to \textit{512*512*2048} bytes in size).
	\item decomposition significantly improves visualization performance, due to the reason that during the rendering of a single small block, the selection comes from a small texture, which is much faster than selecting from a large texture. Thus, blocks are displayed in order from the observer, some of the blocks can be closed by previously drawn blocks. The early beam completion strategy is also applicable to block data rendering. In addition to saving GPU memory, it also saves space that you have to trace with rays. As in the case of saving memory, saving space depends on the choice of block size and, as a rule, the smaller the size, the greater the saving. In the block presentation of data, the strategy of skipping empty areas is applicable with greater efficiency, since the data is divided into small arrays, each of which is easier to fit the bounding box than to the data array as a whole. Blocking provides additional opportunities for paralleling rendering to multiple GPUs, which is especially important for the client-server architecture of the visualizer.
	\item despite the need to mix the rendering results of various blocks in a specific order, each individual block can be rendered independently of the others on any GPU. Computing and data can be distributed across different GPUs. There are works on distributed visualization by the method of volume rendering on cluster systems, which is especially important for scientific visualization, when as a result of numerical modeling huge data arrays are obtained, which cannot be transferred to one local machine.
\end{enumerate}
\paragraph{Cons}:
\begin{enumerate}[(a)]
	\item the main drawback of the block representation is the complexity of the rendering algorithm: when rendering a block, nothing is known about data from neighboring blocks (except for the overlap layer with neighboring blocks); if for each block to provide such access there is a need to access to the remaining blocks that process is able to negate the performance gain. Thus, for example, the implementation of casting shadows and various nonlocal lighting techniques \cite{Hernell} \cite{Hadwiger} \cite{Sattler} \cite{Shanmugam} \cite{Stewart} \cite{Tarini} \cite{Behrens} \cite{Zhukov} \cite{Desgranges} \cite{Hernell:01} \cite{Fairchild}, including techniques requiring the generation of secondary rays \cite{Wyman} \cite{Rezk-Salama} \cite{Weiskopf} \cite{Magnor}, becomes more complicated \cite{Kniss} \cite{Jensen}.
	\item It is worth to highlight the complication of the multi-volume rendering algorithm, in which it is necessary to perform joint rendering of two or more spatial data arrays overlapping in space, each of which has a block representation,
	\item block overlapping means that the voxels on the floors will be duplicated, so if the partition is too small (with block sizes less than 323), the GPU memory savings will be ineffective,
	\item if the block is too small, the total time spent on “switching” between the blocks noticeably increases; after rendering the next block, it is necessary to copy the rendering results from the texture (which was performed on rendering) into the texture from which the reading will be carried out during the rendering of the following blocks. Even when copying only the area of the screen where the block was rendered, the performance already drops significantly when the regular block size is 323 voxels  \cite{Gavrilov}.
\end{enumerate}
The work on slab-based rendering \cite{Mensmann} is very interesting as a study demonstrating not only a new method of slicing blocks along a beam to reduce misses for a ray packet in shared memory, yet also demonstrating inconstancy of gain and its limited amplitude ($ \sim 30 \%$); also shows the difficulty of supplementing the data between the blocks due to the above mentioned pros and cons.
\paragraph{The method of Spheres}
The method of spheres for the analysis of the morphology of complex biological objects in \textit{SVR} values: \textit{Surface-to-Volume Ratio}, \textit{SVR} or \textit{Area-to-Volume Ratio}. It is one of the most important characteristics of all biological objects from the cell to the animal as a whole.
This value characterizes the intensity of  exchange between the biological object as a whole with the environment and has a characteristic dependence on the radius \textit{(R)} of the object as \textit{$ \dfrac{1}{R}$}. With the same measure, you can approach the characterization of the local properties of the object. 
Therefore, it is very important to have a quantitative local characteristic
morphology of the studied object in \textit{SVR} values ($ SVR = \dfrac{S}{V} $) in order to discover, understand and explore the functions performed by its parts and organs. This is especially relevant to a challenge with brain cells such as astrocytes, due to the extraordinary complexity of their shape and relatively poor knowledge\cite{Donchin}.
However, to date has not formed sustainable methodology for calculating local \textit{SVR}s. In addition, there is a problem. Settlements in an acceptable time, as computational complexity of the analysis task local \textit{SVR} is proportional to the square of the number of vertices or triangles $ \sim (O = n^{2})$, and astrocytes reconstructed from electron micrographs (a microscope with a resolution of units of nanometers) have in its composition hundreds of thousands $ n \sim (10^{5}) $ triangles\cite{Gavrilov}.
\paragraph{Surface-to Volume approach}
Let's define a scalar field \textit{SVR} \textit{(X)} in 3D space as following: let \textit{X} is an arbitrary point in space, \textit{G} is the surface of the object under study (in our case it is a closed polygonal surface). Then construct a sphere 
\textit{$\Omega$} of radius \textit{R} with center at point \textit{X}. Let the intersection of \textit{$\Omega$} and \textit{G} be nonzero, then let \textit{S} be the area the surface of an object lying inside the sphere \textit{$\Omega$} (denote this surface by \textit{S}), and \textit{V} as it is the intersection volume of the domains \textit{$ {V_{G}}$} and \textit{$V_{\Omega}$} bounded by \textit{G} and \textit{$\Omega$} (denote this is the locus of points as \textit{$\Theta$}. Then the value of the field \textit{SVR} \textit{(X)} is calculated as (1) and presented on the Figure 3:
\begin{equation}
 SVR(X) =\frac{S}{V} 
\end{equation}
In the case \textit{$ V = 0$}, i.e. if \textit{$ {\Omega  \cap G = 0}$}, then the field at the point \textit{X} is not defined.
\begin{figure}[h]
	\centering
	\includegraphics[scale=0.25,angle=0]{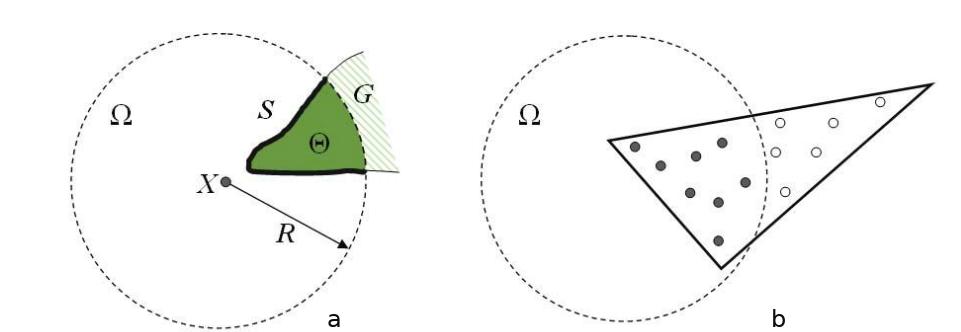}
	\caption{The cross-sectional diagram of the sphere $ \Omega$ for calculating \textit{SVR (X)}: a) bold the part of the surface \textit{G}, the area of which is calculated, the area, volume which is calculated; b) approximate calculation of the area of the triangle inside the sphere $ \Omega$ by the set of random samples of points on a triangle.}
	\centering
\end{figure}
As a result, one defines the method of spheres as the method that consists in calculating the field \textit{SVR (X)} at points \textit{X} belonging to the boundary of the investigated part of the object.
A significant influence on the field values ​​in this method is a choice
the radius of the sphere. To calculate\textit{ SVR (X)}, in accordance with the formulated by definition (1), it is necessary to calculate the area \textit{S} and the volume of only the part \textit{G} located inside the sphere $ \Omega$. Area \textit{S} is the sum of the areas of the triangles located entirely inside $ \Omega$, and from the cut-off part of the areas of triangles intersected by a sphere $ \Omega$. The area of ​the part of the triangle inside $ \Omega$ (see Figure 5 -- b) is calculated by the Monte Carlo method \cite{Sobol}. In a similar way, the volume \textit{G} inside $ \Omega$. However, there are difficulties that should be considered.

\paragraph{Cloud architecture} 
Cloud data storage - an online storage model in which data is stored on numerous servers distributed on the network and provided for use by customers, mainly by a third party. In contrast to the model for storing data on its own dedicated servers purchased or leased specifically for such purposes, the number or any internal structure of the servers is generally not visible to the client. Data is stored and processed in the so-called cloud, which represents, from the user's viewpoint, as one large virtual server. Physically, such servers can be located remotely from each other geographically, up to the location on different continents.
\paragraph{Types of Cloud Computing}
The concept of a  cloud computing is often associated with such service-providing (i.e. everything as a service) technologies, such as:
\begin{enumerate}[(1)]
	\item "Infrastructure as a Service" ("Infrastructure as a Service" or "IaaS"),
	\item "Platform as a Service" ("Platform as a Service", "PaaS"),
	\item "Software as a Service" ("Software as a Service" or "SaaS").
\end{enumerate}
The solution in the study is developing as an open-source solution based on a SaaS model.

SaaS is an application deployment model that involves delivering an application to the end user as an on demand service. Access to such an application is carried out through the network, and most often through an Internet browser. In this case, the main advantage of the SaaS model for the user is the lack of costs associated with installing, updating and maintaining the use of the equipment and software running on it.
In this SaaS model presented on the Figure 6 includes the following points:
\begin{enumerate}[(1)]
	\item the application is studied to be adapted for the remote use,
	\item multiple clients can use one application,
	\item the application can be upgraded by developers smoothly and transparently to users.
\end{enumerate}

\begin{figure}[h]
	\centering
	\includegraphics[scale=0.25,angle=0]{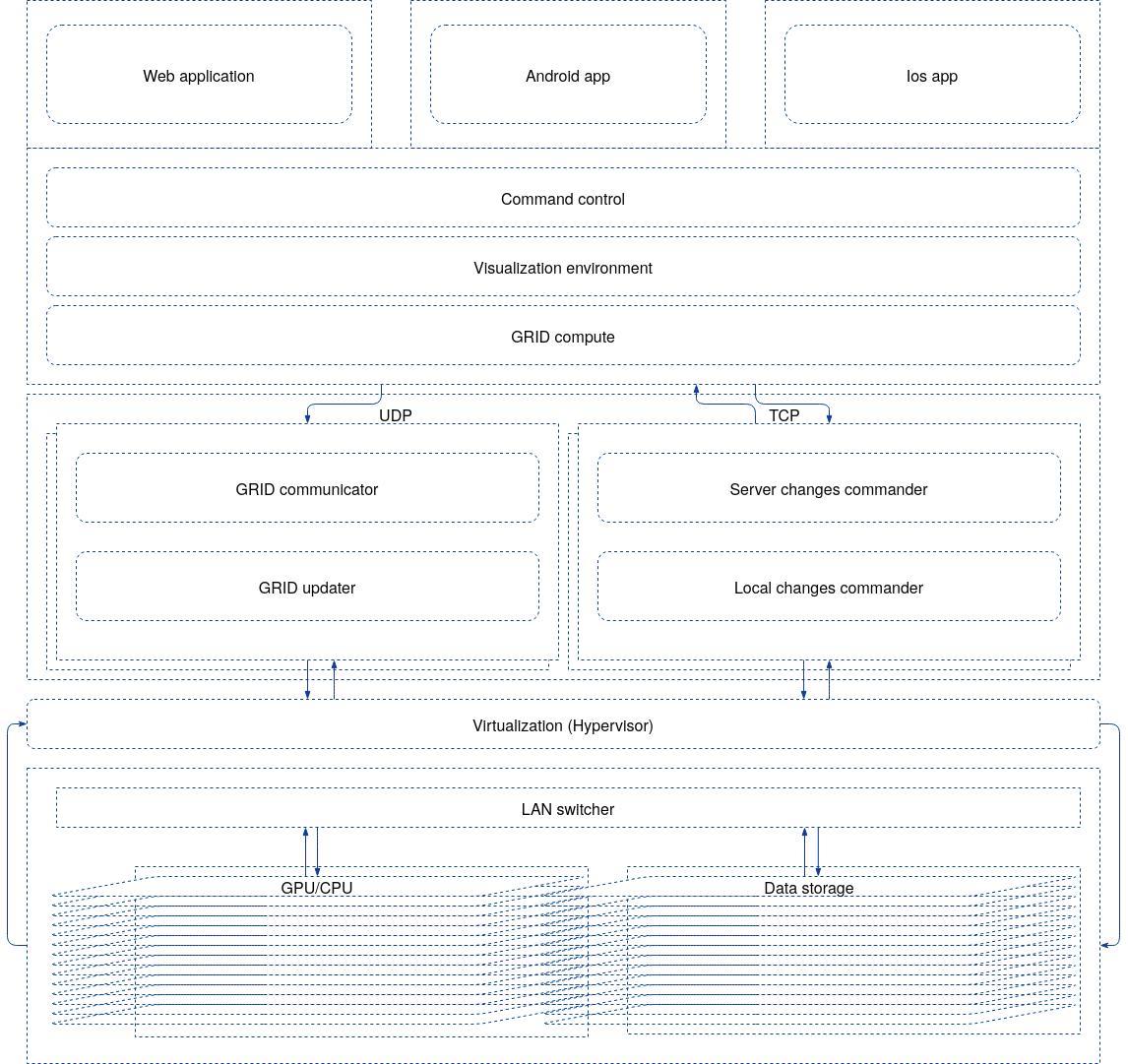}
	\caption{The cloud based server architecture of the SaaS solution}
	\centering
\end{figure}
In fact, SaaS software can be seen as a more convenient and cost-effective alternative to internal information systems.
The development of SaaS logic is the concept of WaaS (Workplace as a Service - workplace as a service). That is, the client receives at his disposal a virtual workstation fully equipped with all the necessary software.
Here on the Figure 6 the approximate cloud architecture is presented.

\section{Results}
There is an arising significant complication of the visualization algorithms coming from the block decomposition of the data. In practice, it's necessity  inevitably arises not only for visualizing large arrays but also to speed up rendering, especially on graphics cards with low performance. 
Regrading this, the method of the block decomposition of data for volume visualization based on the ray emission algorithm (Ray Casting method) is still being studied.
Based on the study, a method for quantifying \textit{DVR} artifacts in a method for ray emission (Ray Casting method) caused by an insufficiently short beam pitch in the form similar to the peaks of the signal-to-noise ratio (PSNR - Peak Signal-to-Noise Ratio) is still being developed. According to the results from the Ray Casting processing in volume render: using the \textit{PSNR} ratio brings the noise level to a logarithmic scale in dB, which values ​​from 30 to 40 dB  corresponding to the  acceptable image synthesis quality. The causes of artifacts arising from the trilinear interpolation are investigated further as much as the cubic interpolation (and either their approaches to the quantitative assessment involving the assessment of structural similarity of two \textit{SSIM} images).      
A new method using the postclassification for eliminating volumetric errors (artifacts) -- visualizations characterized by using pre-integrated rendering in a virtual data sampling method when integrated along the beam, which is optimal for a class of visualization cases where tricubic interpolation and local lighting is proposed, still investigated.
A study was conducted on the 3D visualization methods based on the Ray Casting (RC) algorithm. Although no RC algorithm was found in the course of the experiments that was optimal in terms of quality and performance under any visualization conditions, the evaluation method showed the necessity of optimization: due to the reason that the not optimized \textit{UDVR} approach is inferior to other approaches in any visualization conditions, despite its high performance.

Applying the approaches to volume render of medical image data studied in this work, it was managed to design the 3D visualization software prototype and achieve interactive and high-quality volumetric visualization of medical image of about 2 GB in size (512x512x5382 voxels) on the localhost machines. Also, the design for the cloud based grid machines was developed. The solution mostly implements the hypervisor technologies for the grid server computing.  A method for implementing a fully functional system based on GRID cloud computing servers is proposed, examples of client-side are shown on the Figures 7, 8, 9.

\begin{figure}[h]
	\centering
	\includegraphics[scale=0.25,angle=0]{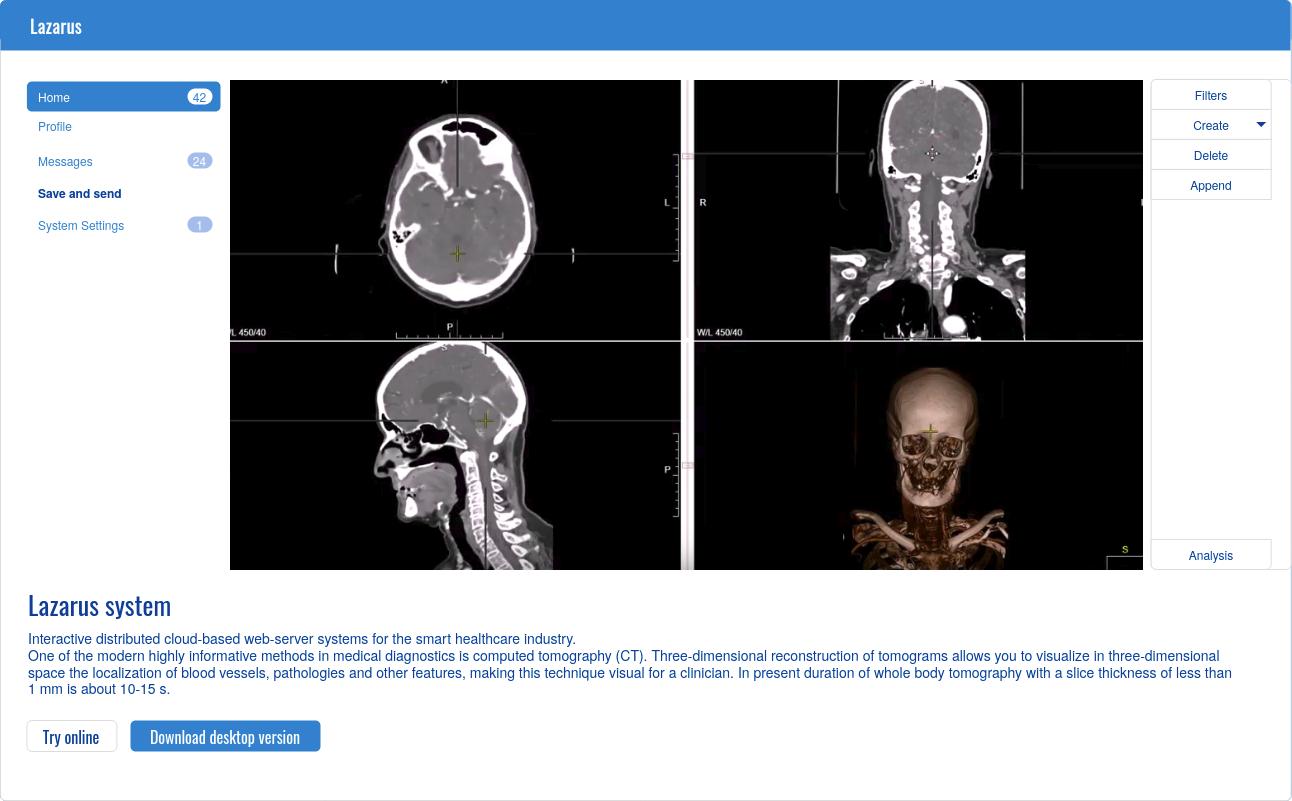}
	\caption{The cloud based online version (prototype)}
	\centering
\end{figure}
\begin{figure}[h]	
	\centering
	\includegraphics[scale=0.25,angle=0]{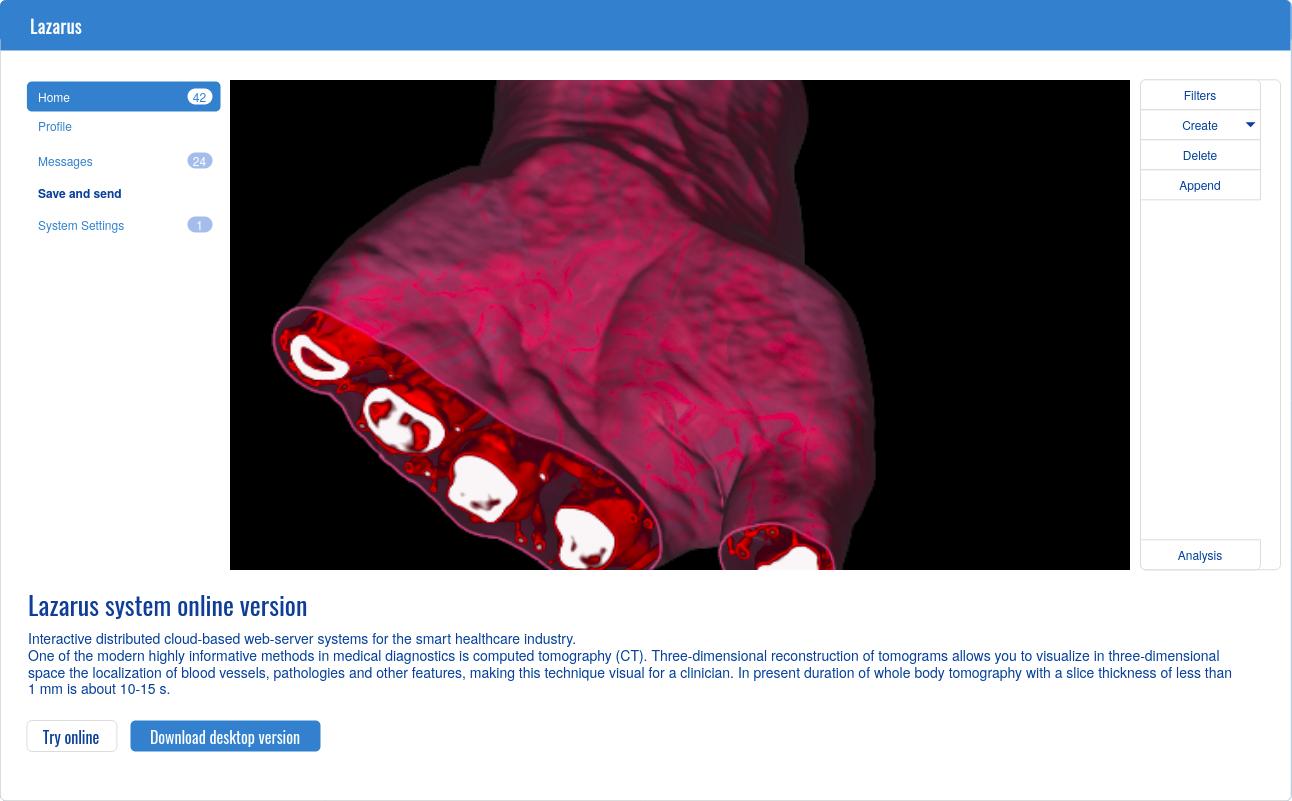}
	\caption{The cloud based online version (prototype)}
	\centering
\end{figure}	
\begin{figure}[h]	
	\centering
	\includegraphics[scale=0.25,angle=0]{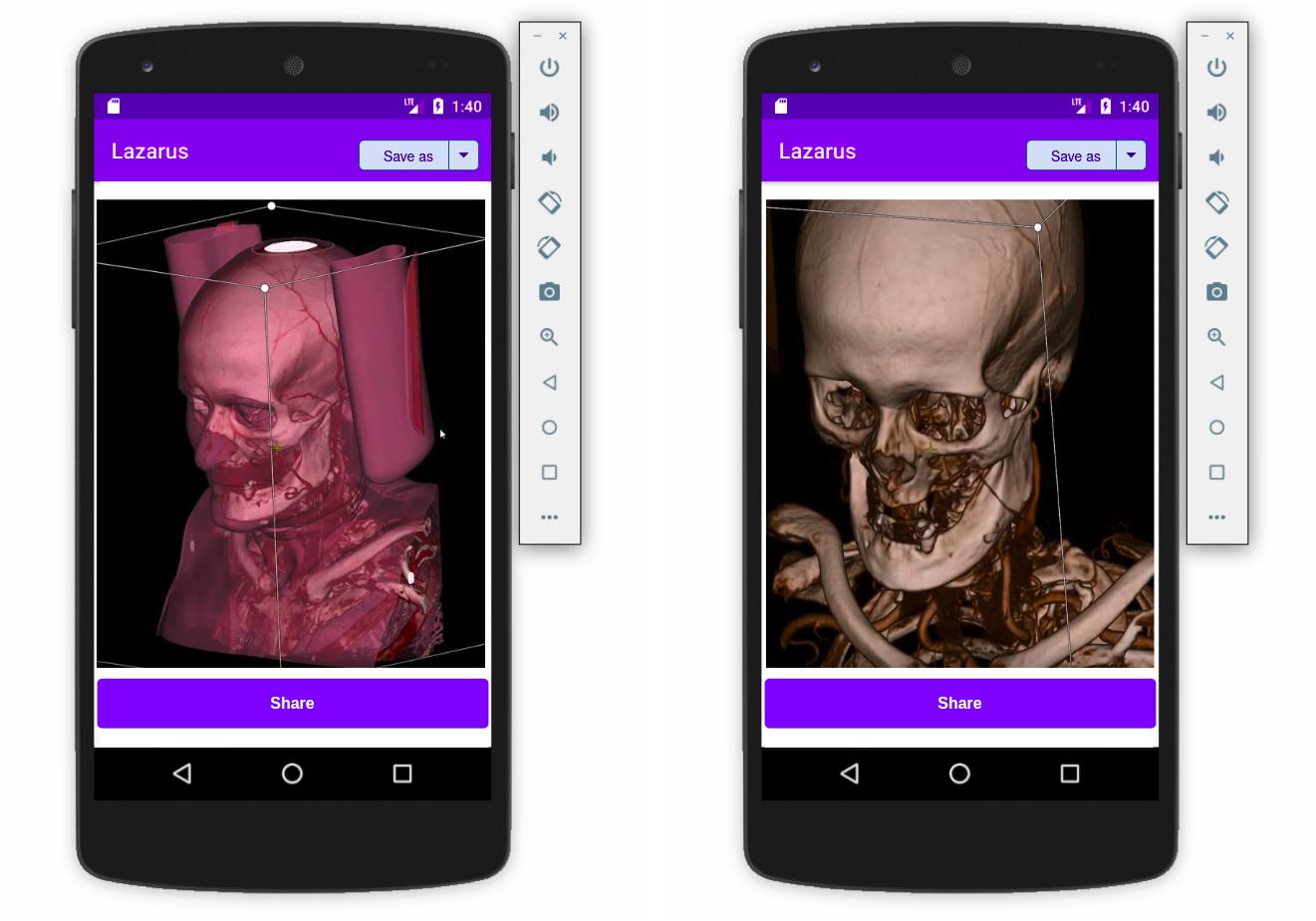}
	\caption{The mobile version (prototype)}
	\centering
\end{figure}

\section{Discussion}
Today there are variety of the approaches developed in the works of the following scientists: Klaus Engel, Bernhard Kainz, Daniel Ruijters, Stefan Guthe, Johanna Beyer, Vincent Vidal, Markus Hadwiger, Daniel Weiskopf, Thomas Ertl, Wolfgang Strasser, Byeonghun Lee, Jihye Yun, Jinwook Seo , Yeong-Gil Shin, Bohyoung Kim, Byonghyo Shim, and others, allowing to process the real-time volumetric visualization using GPU computing. The world market offers several tomography software systems that provide fusion and three-dimensional visualization of tomograms. These commercial systems use the most productive versions of the commercial 3D visualizer systems \cite{Tractica}.
The same can be said about the growth of information flow and the need to build productive methods for processing it in the high-energy physics. The most critical situation, due to the huge amount of data, is observed in three-dimensional processing of data from particle accelerator detectors with a resolution of a few picoseconds ($ \sim10^{-12}$ seconds) \cite{Cheung:1495074}.
For various mathematical models, it is necessary to visualize the obtained data in such a way that certain field properties are revealed. For example, for the result of numerical simulation of unsteady fluid flows, these properties can be revealed:
\begin{enumerate}[(1)]
	\item dynamics of the velocity field,
	\item the formation and decay of vortices,
	\item vortex flows and shock waves.
\end{enumerate}
Of particular difficulty is the task of constructing animations for vector velocity fields (both steady and unsteady flows). Texture methods for solving this problem allow to obtain high-quality results in the visualization of two-dimensional flows. The pinnacle of their development are methods based on the construction of a Motion Map for stationary flows. However, when trying to apply them to the three-dimensional flows, a number of problems arise, associated primarily with the high density of texture data, and also, importantly, with large computational costs.
The bottleneck of the well-known three-dimensional texture methods is the construction of interactive animation with high quality animated paintings. Due to the high computational complexity, it is possible to build only a certain animation sequence for the purpose of its subsequent visualization.
Despite significant progress in solving the problems mentioned above, there are a number of unsolved problems:
\begin{enumerate}[(1)]
	\item high-quality three-dimensional visualization of medical images  today is “tied” to the tomograph due to the high demands on the productivity of the workstation, yet it is not available to the ordinary clinician and, especially, the medical students. Thus, a transition to the online software  for the mass accessibility technology without the loss of visualization quality is necessary,
	\item the volume of the medical image available for the 3D reconstruction on the GPU is limited by the size of the GPU memory (today for the mass office video cards that are for sale the volume is about $ \sim$ 1-2 GB), while the constant growth of data requires the removal of restrictions on their volume and the construction of decomposition algorithms for parallel block data processing while preserving all the capabilities and quality of visualization on the side of user services,
	\item despite the increase in productivity and quality of 3D visualization, there is no practice of quantitatively assessing the quality of visualization; as for the gaming technologies, for example, the so-called light distribution calculation method is used \cite{Magro}
	\item There are several open and commercial programs for semi-automatic three-dimensional geometric reconstruction of cells, but there are no methods and programs for automating detailed morphological analysis of cells, while the computational complexity of such analysis in SVR (Surface-to-Volume Ratio) values is proportional to the square of the number of vertices of triangles \ $O^{2}$\ with characteristic values of approximately\ $\textsl{10*5}$\ triangles and deviation variance of the vortex data given under the volume render \cite{MargretAnouncia:2697932}.
\end{enumerate}

\paragraph{Software rights}
All rights on the software are under the MIT License.

\section*{}
\bibliography{mybibfile}

\end{document}